\documentclass[prd, twocolumn, nofootinbib, floatfix, notitlepage,longbibliography]{revtex4-2}
 
\pdfoutput=1

\usepackage{amsmath}
\usepackage{graphicx}
\usepackage{dcolumn}
\usepackage{bm}
\usepackage{epsfig}
\usepackage{amssymb,latexsym,mathrsfs,empheq}
\usepackage{graphicx}
\graphicspath{{./}{figures/}}
\usepackage{color}
\usepackage{xcolor} 
\usepackage{mathtools} 
\usepackage{hyperref} 
\usepackage{orcidlink}

\hypersetup{
    colorlinks=true,
    linkcolor=red,
    citecolor=blue,
} 

\newcommand{\be}{\begin{equation}}
\newcommand{\ee}{\end{equation}}
\newcommand{\bs}{\begin{split}} 
\newcommand{\bea}{\begin{eqnarray}}
\newcommand{\eea}{\end{eqnarray}}

\newcommand{\omo}{\Omega_{m,0}}

\newcommand{\rhdm}{\rho_{\rm dm}} 
\newcommand{\dldm}{\delta_{\rm dm}} 
 
\newcommand{\omdmo}{\Omega_{{\rm dm},0}} 
\newcommand{\wdm}{w_{\rm dm}}

\newcommand{\fsig}{f\sigma_8} 
\newcommand{\fcl}{F_{\rm cl}}

\usepackage{soul}

\usepackage{breqn}
\makeatletter
\let\cat@comma@active\@empty
\makeatother

\begin{document}

\title{Model Independent Dark Matter Properties from Cosmic Growth} 

\author{Tilek Zhumabek$^{1,2}$\orcidlink{0000-0001-7900-786X}, Mikhail Denissenya$^{1}$\orcidlink{0000-0003-4734-7127}, Eric V.\ Linder$^{1,3}$\orcidlink{0000-0001-5536-9241}} 
\affiliation{
$^{1}$Energetic Cosmos Laboratory, Nazarbayev University, 
Astana 010000, Qazaqstan\\ 
$^{2}$Department of Physics, School of Sciences and Humanities, Nazarbayev University, 
Astana 010000, Qazaqstan\\
$^{3}$Berkeley Center for Cosmological Physics \& Berkeley Lab, 
University of California, Berkeley, CA 94720, USA
} 

\begin{abstract}
Dark matter dominates the matter budget of the universe but its 
nature is unknown. Deviations from the standard model, where dark 
matter clusters with the same gravitational strength as baryons, 
and has the same pressureless equation of state as baryons, can 
be tested by cosmic growth measurements. We take a model independent 
approach, allowing deviations in bins of redshift, and compute the 
constraints enabled by ongoing cosmic structure surveys through 
redshift space distortions and peculiar velocities. These can 
produce constraints at the $3-14\%$ level in four independent 
redshift bins over $z=[0,4]$. 
\end{abstract} 

\date{\today} 

\maketitle

\section{Introduction} 

Dark matter is an essential ingredient of the standard model 
of cosmology \cite{0608407,1807.06209,pdgdm22} and critical to 
explaining the formation of cosmic structure and the properties 
of galaxies \cite{1804.03097}. Indeed, dark matter contributes 
$\sim6$ times more than standard model baryonic matter to the 
energy budget of the universe. However, the properties of dark 
matter, even as it relates to their cosmological effects, are 
not clearly known. 

A plethora of dark matter cosmic properties have been postulated, 
often to address apparent tensions in galaxy structure or 
cosmological parameter determination. These include nonstandard 
interactions, decays, cannibalism, equation of state, etc. 
A few recent examples include 
\cite{2103.01183,2303.06627,2307.08495,2308.13617,2310.06028}. 
One can also take a more phenomenological point of view, often 
called ``generalized dark matter'' \cite{9801234}, where 
an equation of state away from the standard pressureless one 
is allowed, along with a sound speed for the fluid perturbations 
and a possible viscous sound speed. A few recent examples include 
\cite{2209.08102,2307.05155,2307.09522}. 

In this work we focus on what cosmic growth observations can 
say about the two properties that we know dark matter possesses: 
energy density evolution (or equivalently equation of state) 
and clustering strength (i.e.\ gravitational or other). 
That is, how can cosmic data constrain whether dark matter is 
indeed standard: pressureless (equation of state with $\wdm=0$) and 
clustering with gravitational strength ($G=G_N$). In order to 
be as model independent as possible, we will not adopt  
functional forms but rather allow independent values in 
bins of redshift. 

In Section~\ref{sec:pert_eq} we review the growth equation 
of linear, subhorizon density perturbations and identify two areas
in which it may deviate from the standard model. Section~\ref{sec:part_clust} evaluates the case where the gravitational clustering strength of dark matter is modified, while Section~\ref{sec:eos} investigates when the density evolution (equation of state) of dark matter is modified. In each case, we project constraints from the combinations of galaxy redshift surveys. We conclude in Section~\ref{sec:concl}.

\section{Dark Matter and Cosmic Growth} \label{sec:pert_eq} 

Cosmic structure probes not only the expansion rate of 
the universe, including the energy density evolution 
(equation of state) of each component, but the strength with 
which each cluster, either gravitationally or through 
interactions or self interactions. Therefore it is a 
powerful cosmic probe beyond the standard model. Here we will focus on dark matter properties, in a flat universe 
with baryons, dark matter, and a cosmological constant, 
with no conversion between them. 

In the subhorizon, linear density perturbation regime, 
matter grows in the standard cosmological model according to 
\begin{equation}
    \frac{d^2\delta(t)}{dt^2} + 2H(t)\, \frac{d\delta(t)}{dt} -4\pi G_N \rho_m(t)\, \delta(t) = 0\ , 
\end{equation}
where $\delta = \delta \rho_m/\rho_m$ is the matter overdensity, $\rho_m=\rho_b+\rho_{\rm dm}$ is the sum of 
the baryonic matter and dark matter, $t$ is the cosmic time, 
$H = \dot{a}/a$ is the Hubble parameter, and $G_N$ is the Newton's constant. 

One can of course consider cosmologies other than dark 
energy being a cosmological constant, in which case $H$ 
changes (and dark energy perturbations arise, but these 
generally have negligible impact on matter growth). 
Instead we keep the dark energy sector as a cosmological 
constant and explore the dark matter sector, by allowing 
the dark matter clustering strength to deviate from the 
universal gravitational strength, 
\be 
G_N\rho_m\delta \rightarrow G_N\,(\rho_b\delta_b+\fcl\,\rhdm\dldm)\ . 
\ee 
The clustering strength $\fcl$ is treated phenomenologically, 
without a specific model assumed, but time dependence is 
allowed. 
This will give a model independent generalization of 
some investigations in \cite{1302.4754}. 
Such a standard model extension is explored in Section~\ref{sec:part_clust}. 

The second modification considered to the standard model 
is to allow deviations to the dark matter density 
evolution, 
\bea  
\rhdm(a)&=&\rhdm(a=1)\,a^{-3}\notag\\ 
&\rightarrow&\rhdm(a=1)\,a^{-3}\,F_{\rm eos}(a)\ . \label{eq:rhof} 
\eea 
Again we treat $F_{\rm eos}$ phenomenologically. This 
will change $H(t)$ as well as the source term of the 
growth evolution equation. If one altered dark energy as 
well to preserve $H(t)$, this approach becomes equivalent 
to a change $\fcl$ entering just the source term, so we 
keep dark energy fixed as a cosmological constant. 
This extension is explored in Section~\ref{sec:eos}.

In both cases we adopt a model independent approach 
by fitting the deviating functions $F$ in independent bins of 
redshift. We fix the baryons to have standard evolution 
$\rho_b\sim a^{-3}$, with value today 
$\Omega_{b,0}h^2 =0.02233$ indicated by primordial nucleosynthesis and cosmic microwave background (CMB) 
measurements. We also take $h=0.7$ ($\Omega_b=0.0456$) 
and a fiducial $\Omega_{m,0}^{\rm LCDM}=0.3$.

\section{Dark Matter Clustering} \label{sec:part_clust} 

Galaxy redshift surveys map large scale structure over 
a redshift range, and the clustering evolution depends not 
only on the growth factor 
$\delta(t)$ 
but on the growth 
rate $f=d\ln\delta/Hdt=d\ln\delta/d\ln a$ entering into redshift 
space distortions (RSD). It is convenient to work in terms of 
scale factor $a$ or redshift $z=a^{-1}-1$, and also to 
normalize the growth factor to take out the matter dominated 
behavior $\delta^{\rm md}\sim a$ by using 
$g(a)=[\delta(a)/\delta(a_i)]/(a/a_i)$, where $a_i$ is some 
initial condition scale factor in the matter dominated epoch 
(where $g=1$, $f=1$). 

The growth evolution equation can then be written as 
\bea 
0&=&a^2\frac{d^2g}{da^2}+\left(5+\frac{1}{2}\frac{d\ln H^2}{d\ln a}\right)\,a\frac{dg}{da}\\ 
&\quad&+\left(3+\frac{1}{2}\frac{d\ln H^2}{d\ln a} - \frac{3}{2}\left[\Omega_b(a) + F_{cl}(a)\,\Omega_{dm}(a)\right]\right)\,g \ . \notag 
\label{eq:growth}
\eea 

Note that $f=1+d\ln g/d\ln a$ and that RSD  
involves $\fsig(a)$ where $\sigma_8(a)\sim\delta$ is related to 
the mass fluctuation amplitude (so $\fsig\sim d\delta/d\ln a$). 
While one could fix the value of $\sigma_8$ at present as 
measured by mass clustering, we want to keep the high redshift 
universe within the standard cosmological model and so 
normalize in terms of the CMB fluctuation amplitude quantity 
$A_s$. Deviations in clustering strength (or in dark matter 
density evolution in the next section) will therefore alter 
the value of $\sigma_{8,0}$. 
We can write 
\begin{equation}
    f\sigma_8(a) = \frac{\sigma_{8,0}^{\rm LCDM}}{g_0^{\rm LCDM}}\,ag \,\left(1+\frac{d\ln g}{d\ln a}\right)\ . 
    \label{fs8}
\end{equation} 
For our fiducial model, $g_0^{\rm LCDM}=0.779$ and we take 
$\sigma_{8,0}^{\rm LCDM}=0.8$.

\subsection{Sensitivity to Deviations} \label{sec:sens} 

We first investigate how the quantities $g(a)$ and 
$f\sigma_8(a)$ respond to deviations in dark matter 
clustering. For a model independent approach we take 
\be 
\fcl(a)=1+c(a)\ , 
\ee 
with the deviations $c$ given by bins 
in scale factor with amplitude $c_i$. We choose four 
bins with $a = [0.2,0.333], [0.333,0.5], [0.5,0.667], [0.667,1]$. 
For $a<0.2$ we set the deviation to 0, preserving the high 
redshift universe and in particular primordial nucleosynthesis 
and the primary CMB. Then $c_1$ represents a deviation in 
the earliest bin and $c_4$ in the latest bin. 

Since growth 
is a dynamical process we expect the growth at some time to 
be influenced by the conditions at all earlier times. 
Thus a tomographic survey covering a broad range of redshifts 
will be useful in indicating when the deviations arise. 
Note that $\fcl$ can be interpreted as either all the 
dark matter clustering with strength different from 
gravitational ($G_N$), or only some fraction of the dark matter 
clustering, e.g.\ if there are multiple species of dark matter. 

\begin{figure*}[ht]
\centering 
        \includegraphics[width=0.48\textwidth]{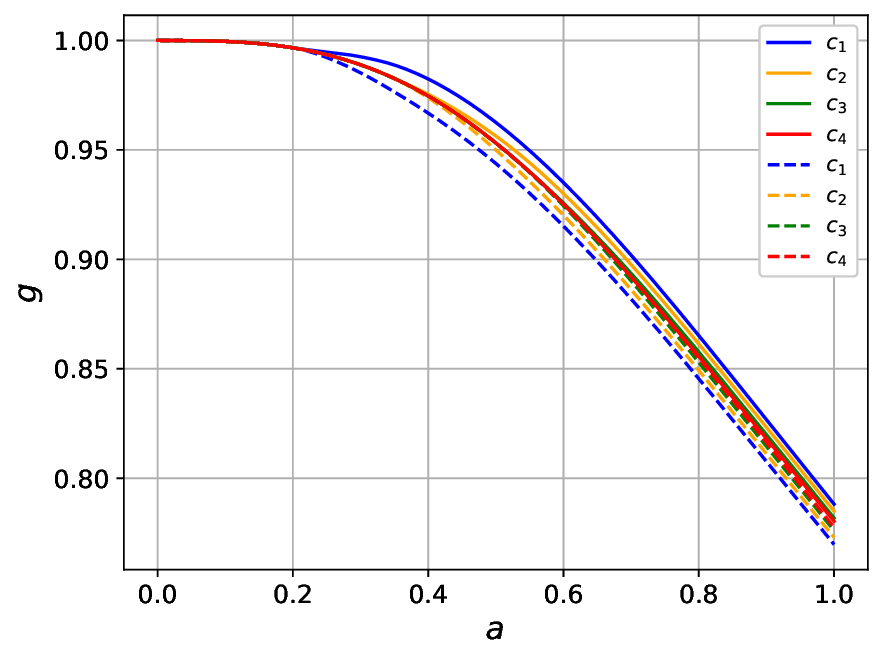}\quad 
        \includegraphics[width=0.48\textwidth]{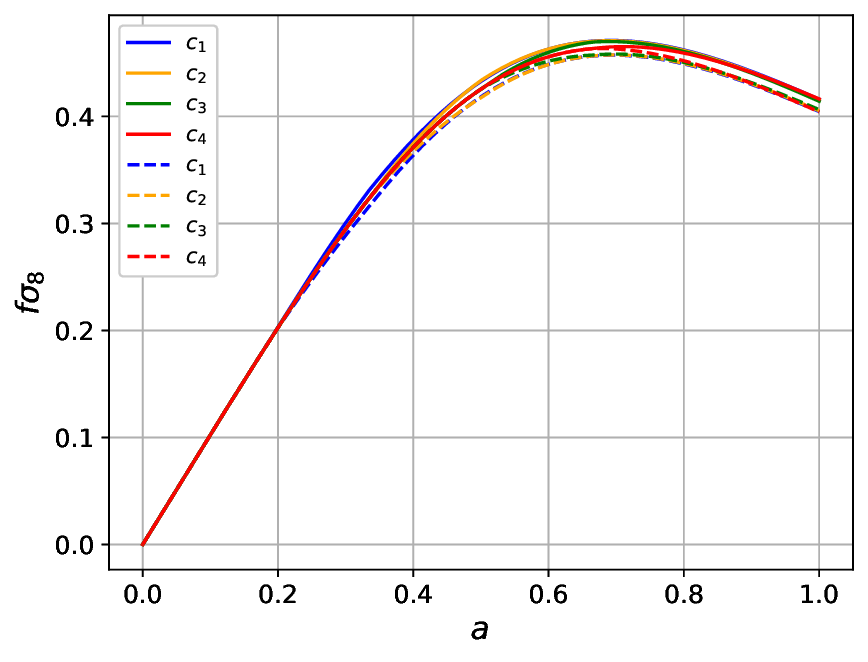} 
\caption{The growth factor $g(a)$ [left panel] and RSD 
factor $\fsig(a)$ [right panel] react to the deviation in 
dark matter clustering behavior. Here each deviation $c_i=\pm0.05$ (solid/dashed curves respectively) is turned 
on one at a time. 
}
    \label{fig:g_fs8}
\end{figure*}

\begin{figure*}[ht]
\centering 
        \includegraphics[width=0.48\textwidth]{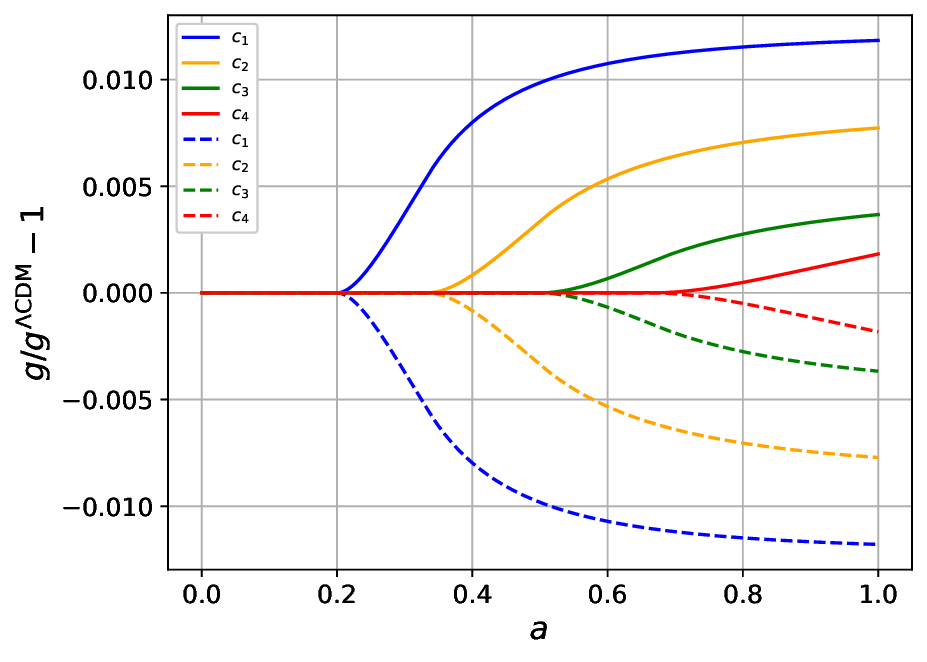}\quad 
        \includegraphics[width=0.48\textwidth]{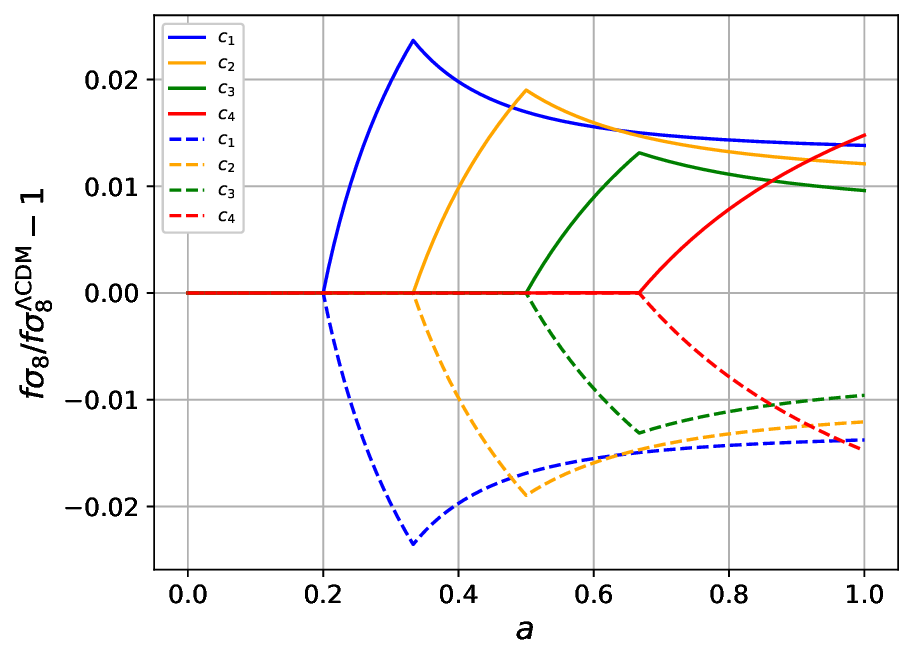} 
\caption{As in Figure~\ref{fig:g_fs8} but shown as residuals 
relative to the LCDM behavior. 
}
    \label{fig:g_fs8_lambda}
\end{figure*}

Figure~\ref{fig:g_fs8} shows the impact of $c_i=\pm0.05$ 
for each bin individually on the growth factor $g$ and 
RSD factor $\fsig$. The fractional deviation of each 
quantity from their LCDM values is given in 
Figure~\ref{fig:g_fs8_lambda}. As expected, a change in 
clustering in the earliest bin has longest, sustained 
impact on the growth. We can enhance or suppress growth 
by changing the clustering from the standard model. 
Note that since we turn on the deviation only for the 
length of a bin, the effect on the growth rate $\fsig$ begins 
to diminish at the end of the appropriate bin, but there is 
an ``inertia'' in $g$ as it is an integral of $\fsig$.

\begin{figure*}[ht]
\centering 
        \includegraphics[width=0.48\textwidth]{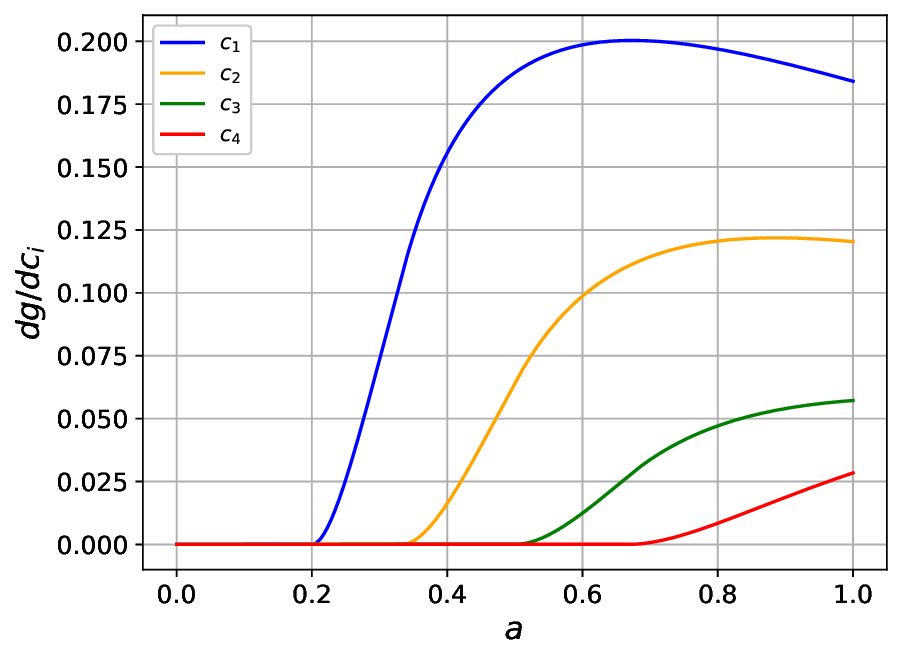}\quad 
        \includegraphics[width=0.48\textwidth]{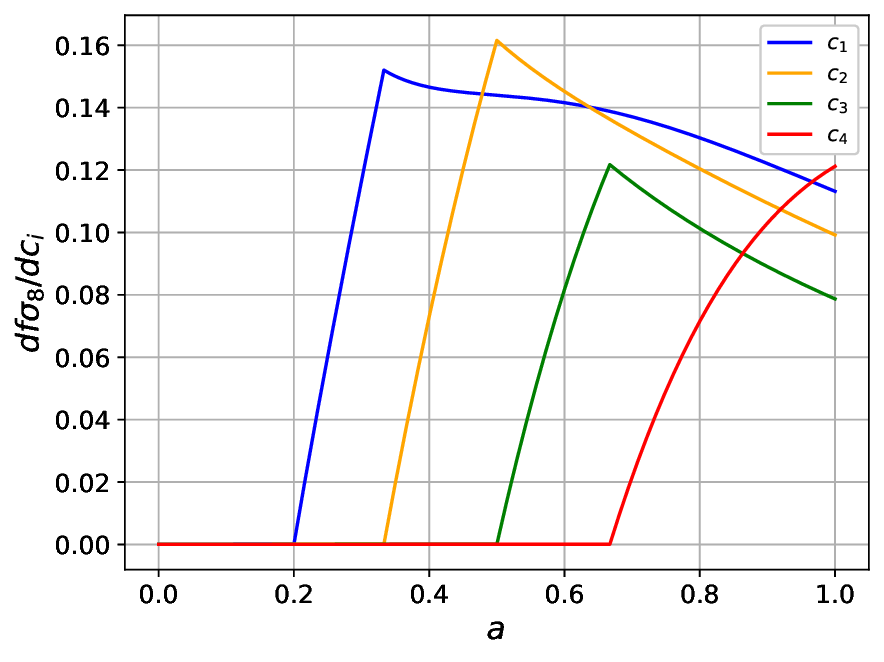} 
\caption{Information sensitivity plots of the growth factor $g$ 
and RSD factor $f\sigma_8$. Since the reactions with respect 
to the clustering strength parameters $c_i$ have different shapes, data over a 
wide range of redshifts can ameliorate covariances between them. 
}
    \label{fig:g_fs8_sens}
\end{figure*}

In order to constrain the deviations with data we will 
employ the information matrix formalism to determine 
$\sigma(c_i)$ for various combinations of surveys. The 
information sensitivity $\partial{\mathcal O}/\partial c_i$ 
of an observable $\mathcal{O}$ is shown in 
Figure~\ref{fig:g_fs8_sens} for $g$ and $\fsig$. We will 
not actually use $g$ in our constraints due to its degeneracy 
with galaxy bias, which can also be redshift dependent. 
This could be lifted using higher order correlation functions 
or data other than growth, e.g.\ weak gravitational lensing 
or the cosmic microwave background, 
but here we only use as data the RSD factor $\fsig$.

Note that the sensitivity curves are distinct in shape, 
which generally indicates that covariances between the 
parameters can be broken by data extending over a broad 
redshift range. The $\fsig$ sensitivity curves rise linearly 
at high redshift, reflecting the results of \cite{1703.00917} 
(see, e.g., Fig.~6 there) that the deviation in $\fsig$ is 
proportional to the area under the curve of the clustering 
deviation, which grows linearly with scale factor between  
the bin boundaries. At more recent scale factor the 
proportionality to area is modified and the sensitivity 
grows somewhat more slowly than linear. After the more 
recent bin boundary is reached, $\fsig$ still has an 
inertia or hysteresis from the deviation and only slowly 
approaches back to the LCDM behavior.

\subsection{Estimated Constraints on Deviations} \label{ref:constrain} 

Given cosmic survey data we can constrain the clustering 
deviations allowed through the standard information matrix 
formalism. As the binned clustering parameters span a 
wide range of scale factor, we consider combinations of 
three different surveys. The mock data will be 2\% precision 
measurements of $\fsig$ from redshift space distortions 
in independent redshift bins 
$z=0.35, 0.45,\dots 1.55$, denoted as ``desi'' since it 
bears similarity to expected measurements from DESI 
\cite{1611.00036}, plus measurements at $z=1.65, 1.75, 
1.85, 1.95$, denoted as ``euclid'' (cf.\ \cite{2020}), 
plus peculiar velocity measurement of $\fsig$ at $z=0.1$, 
denoted ``pv'', as might be enabled by DESI  
\cite{1911.09121,2005.04325,2302.13760}. 
We use lower 
case names to emphasize these are approximate similarities, 
not meant to capture fully the experiments. 
The five fit parameters are $\omo$ and the four $c_i$, with fiducial 
values $\Omega_{m,0} = 0.3$, $c_i=0$. We also add 
a Gaussian prior $\sigma_\omo=0.02$ to represent information 
from external data, e.g.\ supernova and baryon acoustic 
oscillation distances. 

\begin{figure*}[ht]
\centering 
        \includegraphics[width=\textwidth]{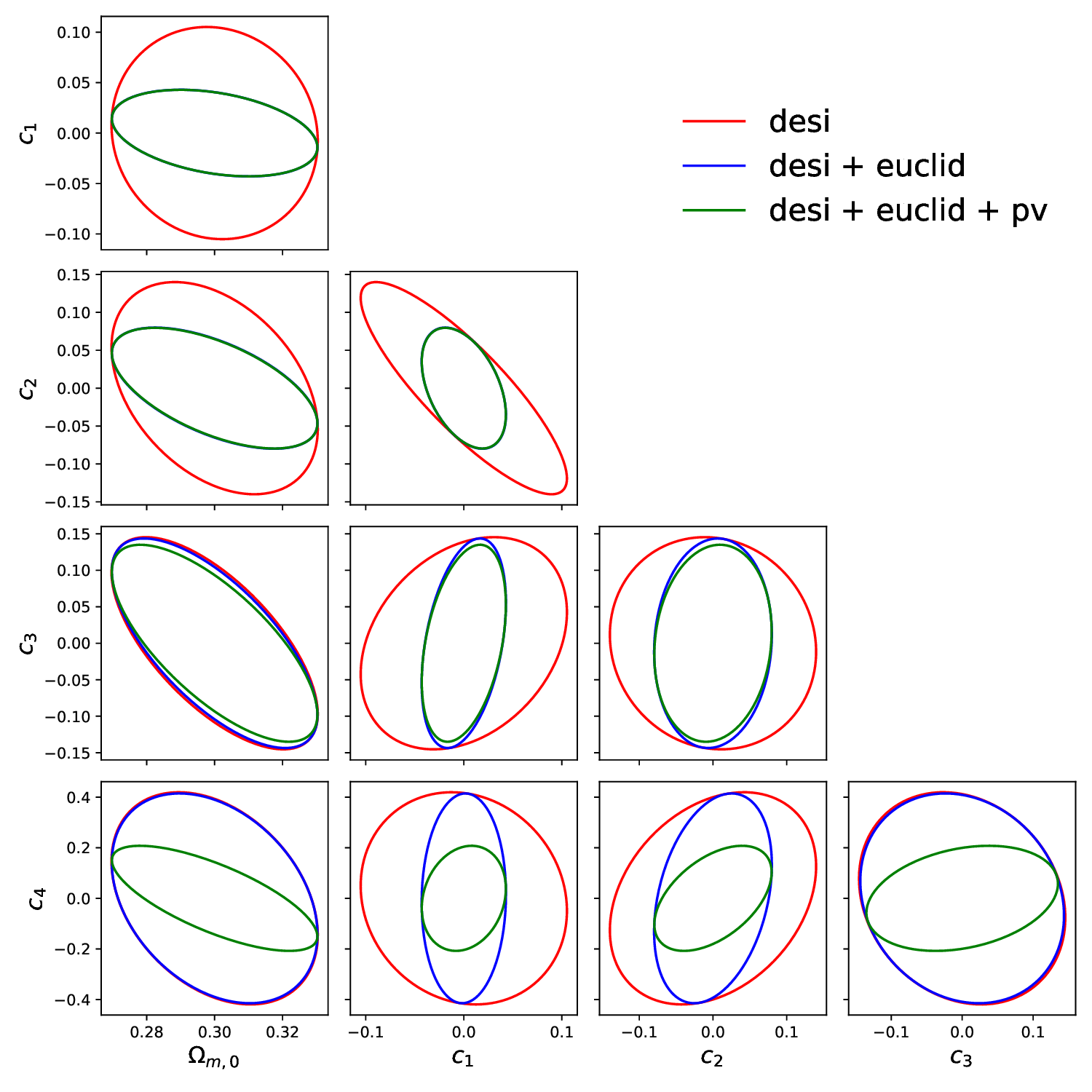} 
\caption{68\% joint confidence level contours on the dark matter clustering parameters and matter density for various data combinations. ``desi'' indicates 2\% precision on $\fsig$ from $z=[0.3,1.6]$, with ``euclid'' adding the range $z=[1.6,2]$, and ``pv'' adding peculiar velocity measurement of $\fsig$ to 2\% at $z=0.1$. 
All cases have a prior $\sigma_\omo=0.02$. 
}
    \label{fig:c_c_combined}
\end{figure*}

Figure~\ref{fig:c_c_combined} illustrates the results 
for three combinations of data. Constraints on $c_1$, in 
the highest redshift bin, are strongly helped by the 
addition of euclid data, while constraints on $c_4$, 
in the lowest redshift bin, benefit from pv data. 
When using all three data sets, the clustering deviations 
are estimated to $\sigma(c_i)=(0.0282,0.0524,0.0888,0.1367)$ respectively. 
The binned values are fairly independent of each other, 
with their strongest correlation being $r(c_2,c_4)= 0.49$.

\section{Dark Matter Equation of State} \label{sec:eos} 

Changing the evolution of the dark matter density influences 
both the source term and the background expansion in the 
growth equation. The density evolution can be phrased in 
terms of the dark energy equation of state parameter $\wdm$. 
We allow this to vary independently in the same four scale 
factor bins as before, with again the standard behavior 
$\wdm=0$ for $a<0.2$. 
The clustering strength is kept at 
the standard gravitational coupling $G_N$ in this section. 

(This basically follows the generalized dark matter 
paradigm of \cite{9801234}, where the dark matter 
properties are described by $\wdm(a)$, the general  
sound speed $c_s(a)$, and the viscous sound speed 
$c_{\rm vis}(a)$. To focus on the influence of changes 
in the dark matter density evolution, we fix $c_s$ and 
$c_{\rm vis}$ to their standard values of zero. In 
any case, 
\cite{0410621,1309.6971,1601.05097,1802.09541,2004.09572} 
demonstrated their effects to be small generally.) 

Since the density evolution is related to the equation of 
state by 
\be 
\rhdm=\rho_{\rm dm}(a=1)\,\,e^{3\int_a^1 (da'/a')\,[1+w(a')]}\ , 
\ee 
then we see that from Eq.~\eqref{eq:rhof} 
\be 
\rhdm(a)=\rho^{\rm LCDM}_{\rm dm}(a=1)\,a^{-3}\,\Pi_{i=1}^j \left( \frac{\min[a,a_{i+1}]}{a_i} \right)^{-3w_i} \ , \label{eq:rhdm} 
\ee 
where the upper limit $j$ denotes the bin in which $a$ is 
evaluated (e.g.\ $j=1$ if $a_1<a<a_2$, and $a_5>1$ so $j=4$ 
for $a_4<a\le1$). For $a<a_1$ then the product is taken to 
be 1, and so $\rhdm(a<a_1)=\rho^{\rm LCDM}_{\rm dm}(a<a_1)$, 
i.e.\ the high redshift behavior is unaffected, as desired. 
Since we keep $h$ (i.e.\ $H_0$) fixed, then 
\be 
\omdmo=\Omega^{\rm LCDM}_{dm,0}\,a_1^{3w_1} a_2^{3(w_2-w_1)} a_3^{3(w_3-w_2)} a_4^{3(w_4-w_3)}\ . \label{eq:omdmo} 
\ee 
As usual $\omo=\Omega_{b,0}+\omdmo$; note that when $w_i\ne0$ 
then $\omo\ne\Omega^{\rm LCDM}_{m,0}$. When all $w_i=0$ 
then the expressions restore to the standard cosmology. 

\begin{figure*}[ht]
\centering 
        \includegraphics[width=0.48\textwidth]{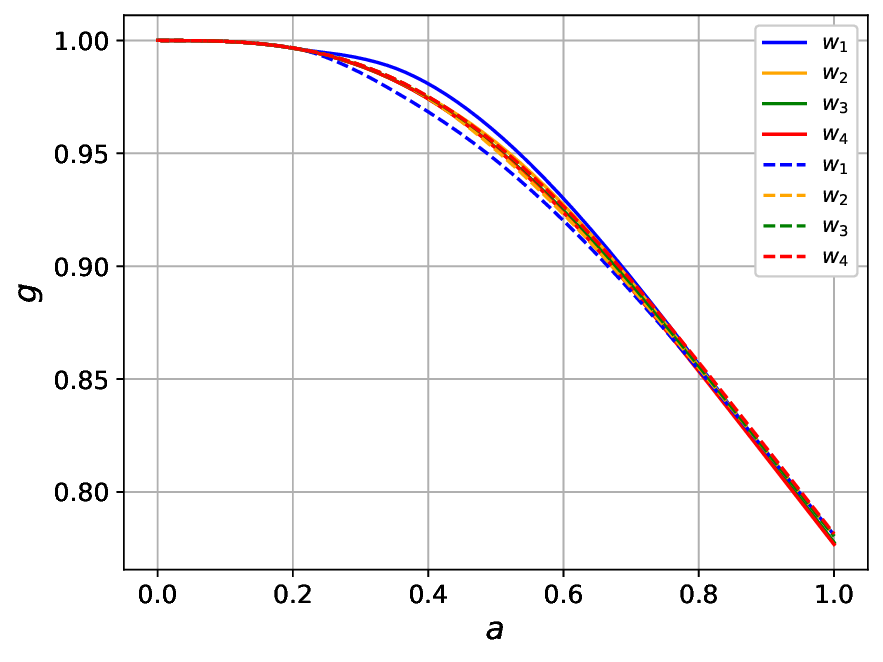}\quad 
        \includegraphics[width=0.48\textwidth]{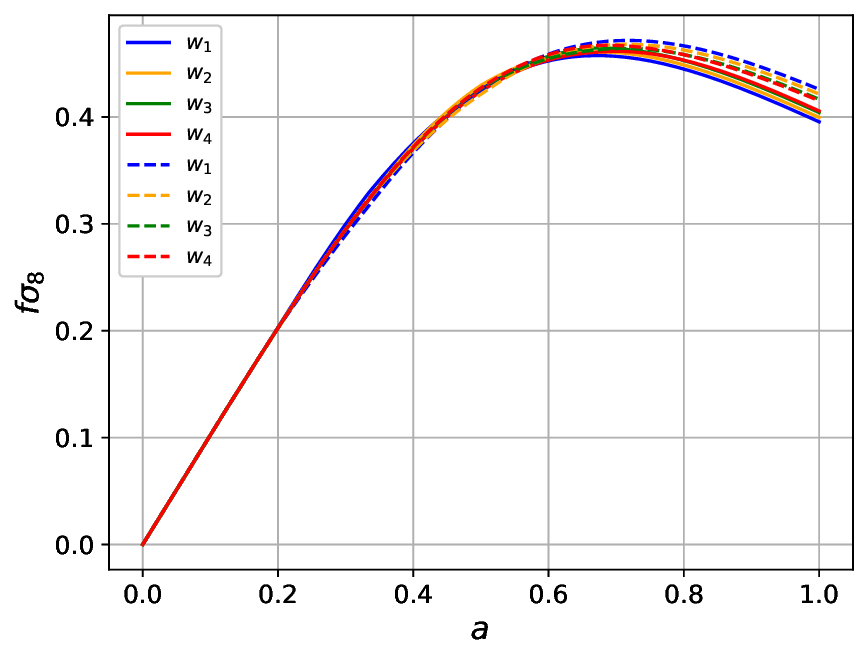} 
\caption{
The growth factor $g(a)$ [left panel] and RSD factor $\fsig(a)$ [right panel] react to the deviation in dark matter equation 
of state. Here each deviation $w_i=\pm0.05$ (solid/dashed curves respectively) is turned on one at a time.
}
    \label{fig:wgfs8}
\end{figure*}

\begin{figure*}[ht]
\centering 
        \includegraphics[width=0.48\textwidth]{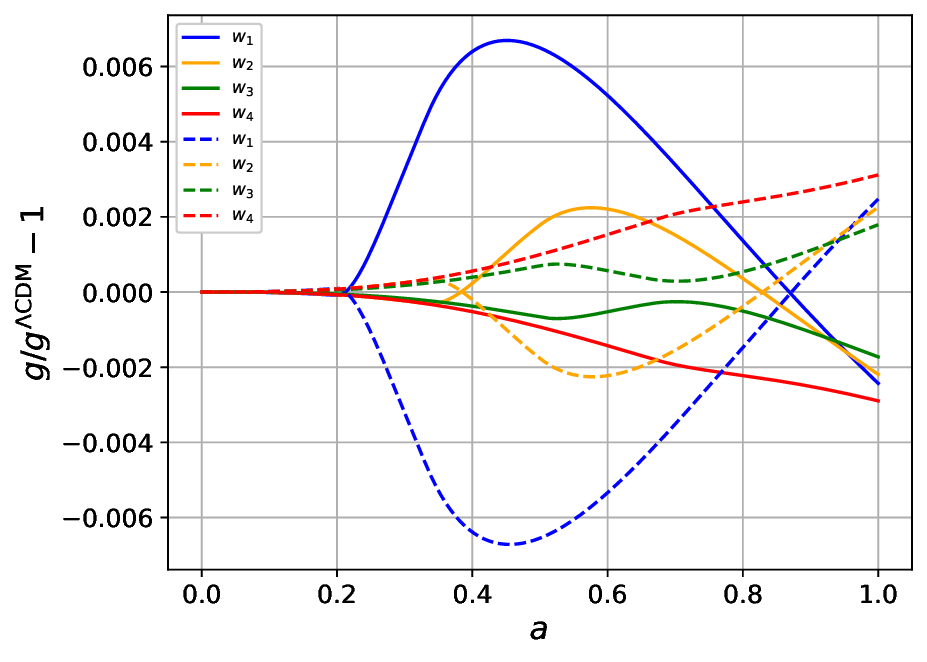}\quad 
        \includegraphics[width=0.48\textwidth]{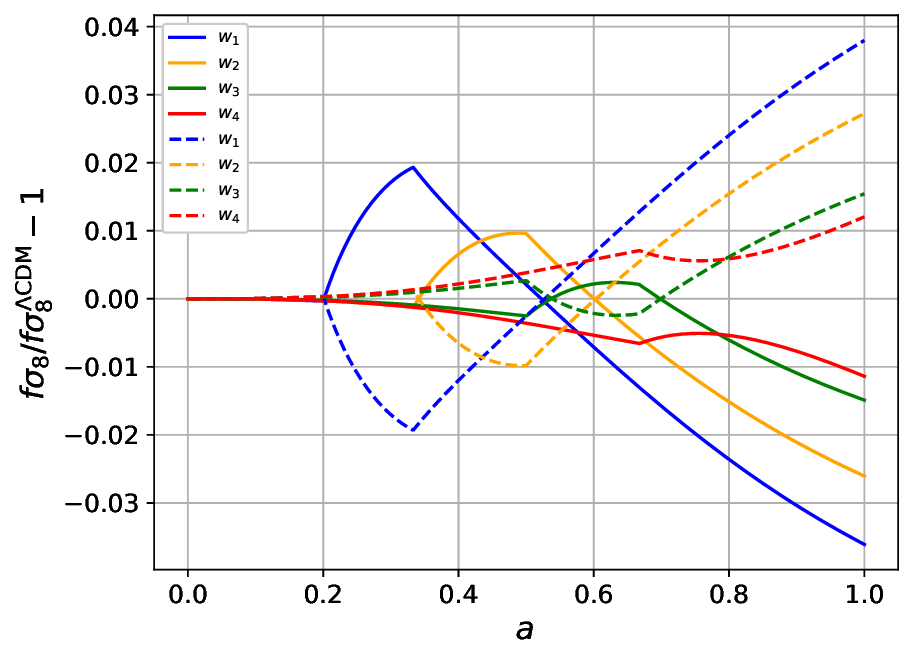} 
\caption{As in Figure~\ref{fig:wgfs8} but shown as residuals relative to the fiducial LCDM behavior.} 
    \label{fig:wresid}
\end{figure*}

The growth factor $g$ and RSD factor $\fsig$ are plotted  
in Figure~\ref{fig:wgfs8} for individual deviations 
$w_i=\pm0.05$. We note that there is actually a crossing 
of the LCDM behavior in each curve. This is clearer in 
Figure~\ref{fig:wresid} that shows the 
residuals with respect to LCDM. 
Here there are three stages to the deviations in growth. 
Although in the matter dominated epoch before the first 
redshift bin for equation of state deviation $w_i$, the 
model is LCDM, it is not the {\it same\/} LCDM model as 
the fiducial because $\omdmo\ne\Omega^{\rm LCDM}_{dm,0}$ 
as seen from Eq.~\eqref{eq:omdmo}. In particular, at 
high redshift before the first bin, Eqs.~\eqref{eq:rhdm} 
and \eqref{eq:omdmo} give 
\be 
H^2(a<a_1)/H_0^2=\Omega^{\rm LCDM}_{m,0} a^{-3}+1-\Omega_{b,0}-\omdmo\ , 
\ee 
so the cosmological constant term is offset. 
Only when all $w_i=0$ 
does $\omdmo$ and $H^2(a<a_1)$ fully restore to the fiducial.

Since $H^2(a<a_1)$ differs from the fiducial LCDM, as $a^3$
(so by very little at high redshift), the growth will actually 
be affected by a small amount even before the first bin. 
From Eq.~(7) in \cite{0701317}, it is straightforward to calculate 
that this implies that the growth rate $f$ (and hence the 
growth factor $g$) deviates from the fiducial as $a^3$. 
Since 
$\omdmo$ decreases as $w_i$ increases, this suppresses growth 
for positive $w_i$. This defines the first stage (pre-bin) in 
the fractional deviations shown in Figure~\ref{fig:wresid}. 
The next stage occurs within the bin where $w_i$ deviates from 
zero. Here, the friction term in the growth equation is 
suppressed (enhanced) and the growth increases (decreases) 
for positive (negative) $w_i$. Recall that 
$d\ln H^2/d\ln a=-3(1+w_{\rm tot})$ and increasing the 
dark matter equation of state parameter increases the 
total equation of state $w_{\rm tot}$. Finally, after the 
bin deviation is turned off, again the reduced $\omdmo$ 
suppresses growth, though more modestly than at the $a^3$ 
rate in the matter dominated era. 

\begin{figure*}[ht]
\centering 
        \includegraphics[width=0.48\textwidth]{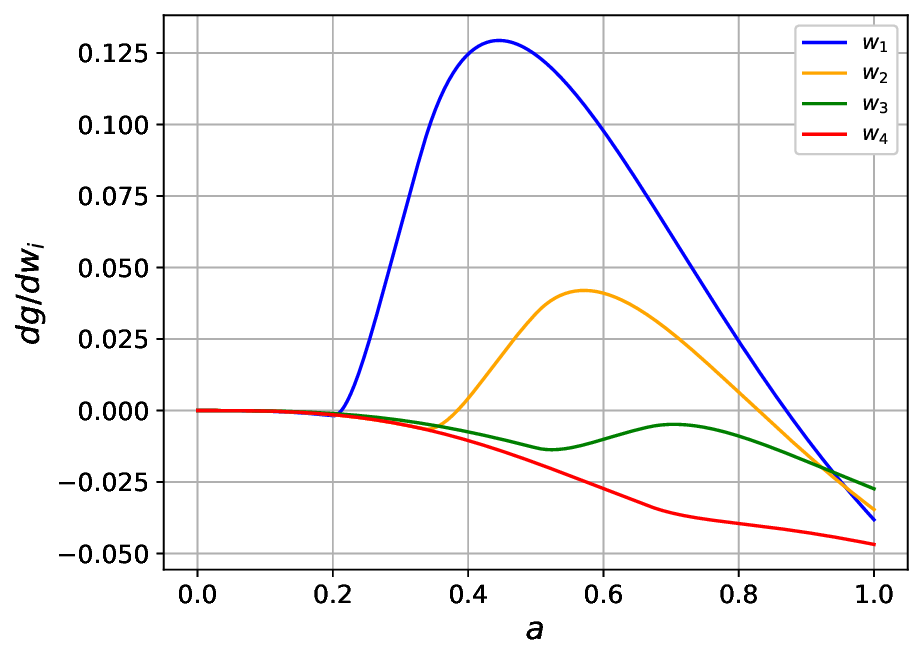}\quad 
        \includegraphics[width=0.48\textwidth]{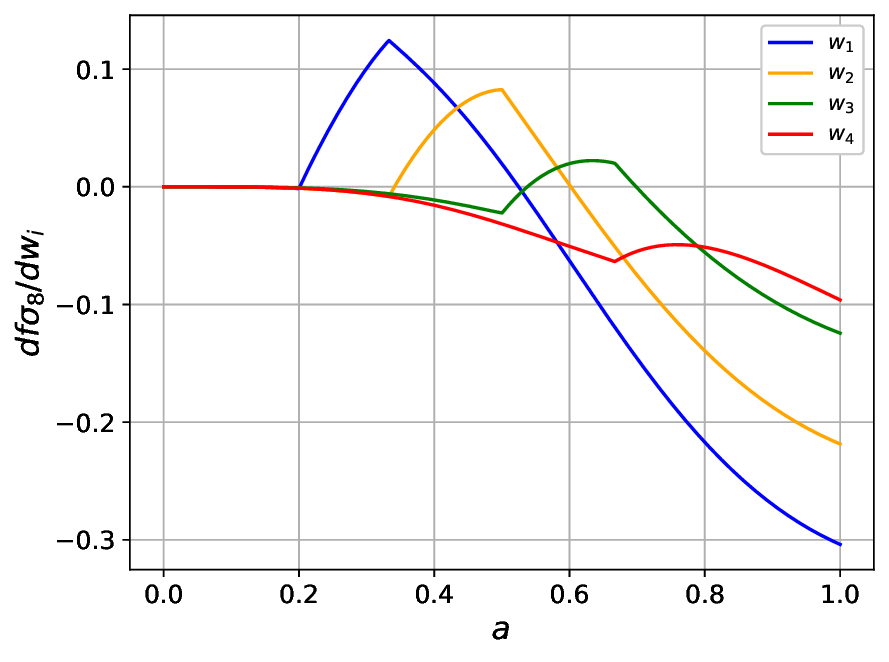} 
\caption{Sensitivity of growth factor $g$ [left panel] and RSD factor $f\sigma_8$ [right panel] to the dark matter equation 
of state parameters $w_i$.}
    \label{fig:g_fs8_sensw}
\end{figure*}

The sensitivity derivatives plotted in 
Figure~\ref{fig:g_fs8_sensw} again show these three 
physical behaviors. However now, 
as with the 
clustering deviations, an equation of state deviation 
in a bin has an inertial effect on the growth, persisting 
somewhat after the deviation turns off until the reduced 
$\omdmo$ takes over.

The fit parameters in this section are 
$\Omega^{\rm LCDM}_{m,0}$ and the four $w_i$, with 
fiducial values $\Omega^{\rm LCDM}_{m,0}=0.3$ and $w_i=0$. 
We expect somewhat more covariance among the parameters 
than previously due to their appearance in $\omdmo$, 
Eq.~\eqref{eq:omdmo}.

\begin{figure*}[ht]
\centering 
\includegraphics[width=\textwidth]{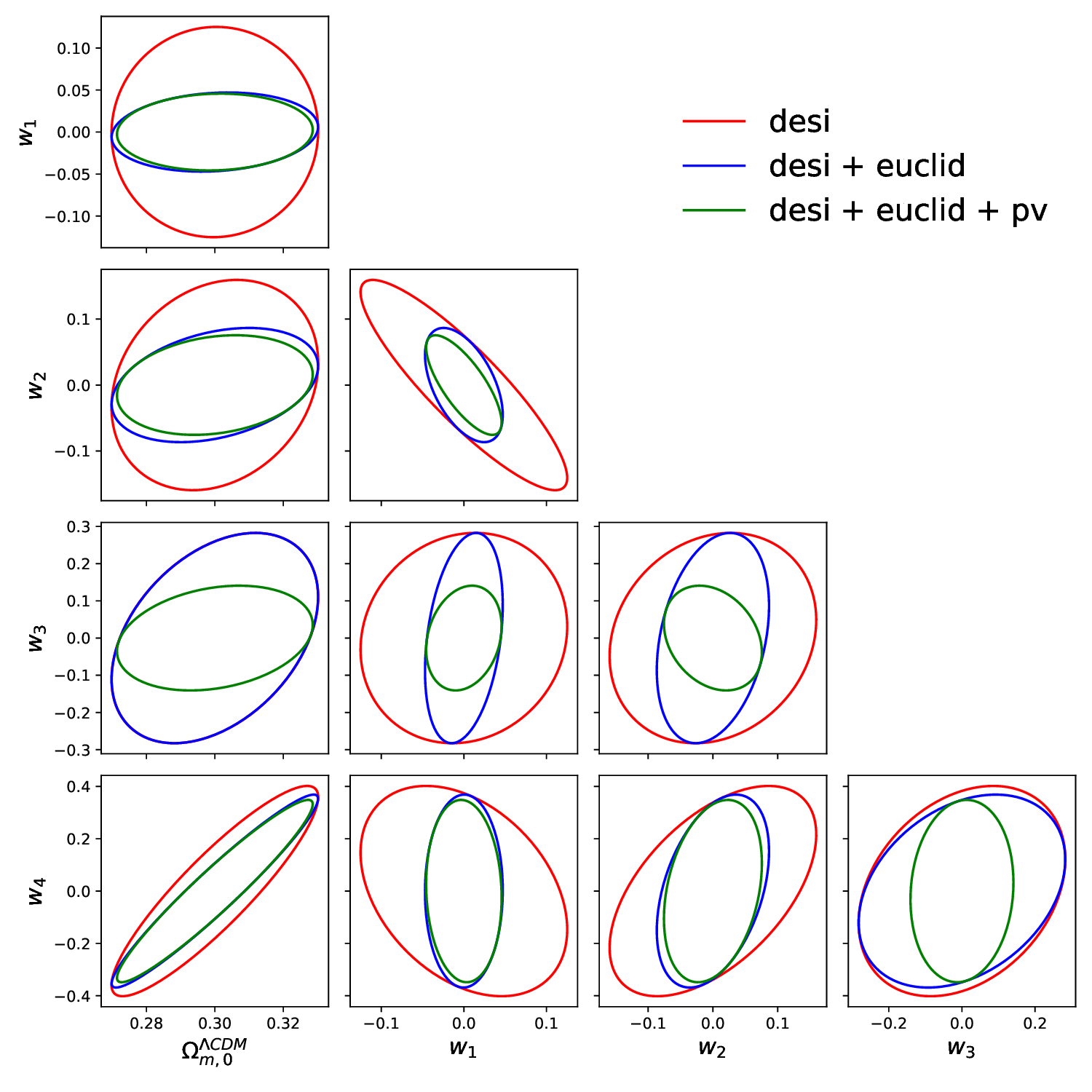}
\caption{68\% joint confidence level contours on the dark matter equation of state parameters and matter density for various 
data combinations. All cases have a prior $\sigma_\omo = 0.02$. 
}
    \label{fig:cornerw}
\end{figure*}

Using the same combinations of data sets as previously, 
Figure~\ref{fig:cornerw} presents the 2D joint confidence 
contours for the parameter constraints. 
When using all three data sets, with the matter density 
prior, the dark matter equation of state deviations 
are limited to 
$\sigma(w_i)=(0.0300, 0.0497, 0.0926, 0.2293)$  
respectively. 
The binned values are fairly independent of each other, 
with their strongest correlation being $r(w_1,w_2)= -0.75$. 

Note that the impact of the low redshift peculiar velocity 
measurement is weaker in this case, especially on $w_4$. 
This is due to two factors: as seen in 
Figure~\ref{fig:g_fs8_sensw} the sensitivity of $\fsig$ 
at low redshift directly to $w_4$ is smaller than to 
the other $w_i$ (compare Fig.~\ref{fig:g_fs8_sens}), 
and the covariance between $w_4$ and $\omdmo$ is much 
higher, with correlation coefficient $r(w_4,\omdmo)=0.95$, 
reducing the constraining 
power of the low redshift measurement. 
The peculiar velocity measurement does help 
with the next lower bin, $w_3$. Adding high redshift 
measurements improves the higher redshift bins, i.e.\ 
constraints on $w_1$ and $w_2$.

\section{Conclusions} \label{sec:concl} 

Dark matter is prevalent on cosmic scales but apart from 
its density we have relatively little idea about its 
properties. Here we examined two characteristics -- its 
clustering strength and equation of state. Our approach 
emphasized model independence, with independent bins in 
redshift. 

We first test whether dark matter clusters with gravitational 
strength. Cosmic growth of structure is an excellent probe 
for this, and we find a strong complementarity between 
surveys covering a diversity of redshift ranges.  
Redshift space distortion measurements of a precision 
similar to Euclid and DESI, and low redshift peculiar velocity 
measurement as could be enabled by DESI and other experiments, 
can constrain the clustering strength at the 3--14\% 
level over the four independent redshift bins. This could 
be tightened by using fewer bins to look for a deviation 
signal, then fine tuning the bins to characterize it. 

The second investigation considers constraints on deviation 
of the dark matter equation of state from zero, 
i.e.\ $\rho_{\rm dm}\nsim a^{-3}$. Here 
the constraints on the four binned values are at the 
$\sigma(w_i)=0.03-0.23$ level. 
This limits how much dark matter can deviate from 
pressureless behavior. 

One can of course obtain tighter 
constraints in both cases by adopting a specific functional 
form for the deviation behavior, but given the uncertainty 
about the dark matter sector we prefer the model independent 
approach. 
Future surveys to even higher redshift, such as 
MegaMapper \cite{2209.04322} or Spec-S5 \cite{2209.03585}, 
could also improve constraints, adding to their science 
cases for dark energy \cite{2106.09581, 2106.09713}. 
Further dark matter properties such as its 
(nonadiabatic) sound speed and viscous sound speed could 
be constrained in a similar model independent manner; we 
leave this for future work.

\acknowledgments 

This work was supported in part by the Energetic Cosmos 
Laboratory. EL is supported in part by the U.S.\ 
Department of Energy, Office of Science, Office of High Energy 
Physics, under contract number DE-AC02-05CH11231.

\bibliography{DM_cl}

\end{document}